\documentclass{article}

\usepackage[english]{babel}
\usepackage{eurosym}

\usepackage[usenames]{color}

\begin{document}

\title{Personal Multi-threading}


\author{%
	Jan A. Bergstra\\
  {\small Informatics Institute, Faculty of Science,}\\
   {\small University of Amsterdam}%
	  \thanks{Author's email address: {\tt j.a.bergstra@uva.nl}, and institutional homepage: 
	  {\tt https://staff.fnwi.uva.nl/j.a.bergstra}.}
\date{}
 }   

\maketitle

\begin{abstract}
\noindent 
Multi-threading allows agents to pursue a heterogeneous collection of tasks in an 
orderly manner. The view of
multi-threading that emerges from thread algebra is applied to the case where a single agent, who may be
human, maintains a hierarchical multithread as an architecture of its own activities.
\end{abstract}

\tableofcontents

\section{Introduction, PMTh}\label{sec:Intro}
Multi-threading drawn from computing proves an attractive metaphor, or perhaps even a model, 
for the simultaneous effectuation
of different tasks or of bundles of tasks for any agent, including human agents. Multi-threading in turn requires a
theoretical model in computing where several accounts of multi-threading have been proposed. Most accounts of 
multi-threading are derived from traditional practices of operating system design (see for instance~\cite{Mueller1993,Barney2009}). Theoretical and conceptual accounts devoted to multi-threading  as a primary subject are scarce.

That motivated the development of the thread algebra model of threads and multi-threads which has 
been proposed and worked out in considerable detail  in \cite{BergstraL2002,BergstraM2006,BergstraM2007,BergstraM2007b,BergstraM2008, BergstraM2010}, and \cite{BergstraM2011}. I will take thread algebra as the point of departure for ``personal multi-threading'' (or single agent multi-threading if one prefers not to highlight the ``human'' aspect). The justification for doing so 
is based on a number of properties of thread algebra, as listed in~\ref{TAproperties} below, which make it deviate from other theories commonly used for the explanation of threads.\footnote{%
In the literature on threads in the tradition of operating systems  a distinction between threads and processes if often made, threads being more lightweight. In theoretical accounts, however, that same difference is 
usually not made and both threads and processes are understood as
instances of concurrent processes to which one of the existing process theories is declared equally applicable. The thread algebra approach deviates from that tradition by suggesting strategic interleaving for threads where process theories, such as e.g. the process algebra muCRL2 documented in~\cite{GrooteM2014} use 
arbitrary interleaving to explain concurrency.}

This paper will be devoted to the special case where 
the activities of a single human agent are understood in terms of multi-threading. I will speak of personal multi-threading, which will be abbreviated to PMTh throughout the paper.

\subsection{Concurrency, multi-tasking, multi-threading, arbitrary interleaving, and strategic interleaving}
A central difficulty in writing this paper, which unfortunately I have not been able to solve in a fully 
satisfactory manner, is to provide and subsequently  use clear distinctions between a range of concepts concerning parallel processes and systems. I will not distinguish between parallelism and concurrency. Concurrent processes occur for instance in an organization if two of its agents
work independently and more or less simultaneously on different and disjoint tasks. Arbitrary interleaving versus true concurrency concerns ways of theorising about concurrent systems, not the mechanics of such systems per se.
The concept of concurrent processes underlies each intuition of parallelism and cannot, in general, be replaced by any form of multi-threading. Concurrency, however, viewed as a behavioural concept extends well beyond concurrent processes. 
Concurrency also includes sequential processes which might be more easily understood as parallel one's though in fact, that is in a mechanical sense, they aren't.

Concurrent processes need not be independent. Many forms of interaction between concurrent processes 
have been proposed and used in computing. I mention synchronous communication, asynchronous communication, communication via shared data, observation of signals, and the imposition of relative priorities between processes 
and between actions or classes of actions. In the present paper concurrent processes will only
play a secondary role to thread.

My present focus wil be on single agents. For single agents concurrency applies via the mechanisms of multi-threading and multi-tasking only. In the psychological literature I could not find a clear distinction between both mechanisms and I run the risk of making design choices concerning the meaning of both which lack support from well-established sources. Nevertheless some choices must be made if the full spectrum of intuitions about concurrency 
is to be made available for the topic at hand. 

I will assume that for a single agent multi-threading refers to the alternating
effectuation of actions belonging to different threads. Threads group actions together which are related by 
a common purpose, objective, goal, cause, or other form of logic. 
The alternating choice of actions from various threads
is referred to as an interleaving (of the behaviour of) the mentioned threads. 
If premeditated sequences of alternation
are put into effect I will speak of strategic interleaving. If, however, some non-deterministic, perhaps randomized,  perhaps partially context dependent, mechanism is governing which thread is requested or permitted 
to carry out its next action, I will speak of arbitrary interleaving. If threads are placed in a sequence and new threads are only initiated after the previous  thread has been put into effect till completion, I will speak of poly-threading
(see~\cite{BergstraM2011}). Poly-threading is an extreme case of strategic interleaving if the next thread is determined by some possibly interactive strategy, while it is an extreme case of arbitrary interleaving otherwise.

Multi-tasking refers to a single agent performing different progressions of actions (together constituting or rather implementing a task) simultaneously, or in overlapping time intervals. An example is agent human $A$ who may walk (task one) and speak (task two) at the same time. $A$'s speaking may be structured in 
several threads.\footnote{%
A conversation between two agents $A$ and $B$ which is made up from two threads, each dealing with different topics, constitutes a seemingly obvious example of multi-threading from a psychological point of view (an example
of a description of this situation can be found in~\cite{PopolovCL1998}). From the 
perspective of theoretical informatics the example is remarkably complex: both $A$ and $B$ produce multi-threaded 
(in fact 2-threaded, or if executive threads are involved 3-threaded) behaviours which unfold concurrently. 
$A$ and $B$ maintain synchronous communication along different channels, one for each thread in the original description. If one takes into account that orderly communication requires that (the processes representing $A$ and $B$) need different names for each spoken thread and that such names must be dynamically created, shared between $A$ and $B$, and subsequently released after thread termination, it follows that the mechanisms of mobile processes from \cite{MilnerPW1992} enter the picture for this seemingly 
very simple example. Further it may be noticed that if $A$ and $B$ were 
chatting instead of speaking 
they would make use of asynchronous communication along different channels  instead.} 
As a part of the walking $A$ may include some thread involving physical exercises.

A single task may be performed in a multi-threaded manner, that is a task may be a multi-thread, and a thread may consist of a multi-task. This gives rise to a layered model. At top level I assume the presence of an overall a multi-thread, this is an arbitrary choice as the thread might consist of a single thread consisting of a multi-task.

\subsection{Multi-threading in psychology versus multi-threading in computing}
Both multi-threading and multi-tasking have been amply studied in psychological research. Nevertheless, it is difficult to avoid inconsistencies between the terminology that I am proposing below and seemingly similar terminology in the psychological literature. For instance in~\cite{SalvucciT2008} it is proposed to view cognitive multi-threading as a mechanism for controlling concurrent multi-tasking whereas I will assume that control of a task is part of that task, thus not making that particular differentiation between task and thread, with a thread being at a cognitive level in charge of a task.

In~\cite{BaiJMD2014} multi-threading occurs within a single task in a multi-tasked setting, in fact
multi-threading is merely a description for a scene which a person must deal with during one of the tasks. 
This use of these terms is inconsistent with mine. This paper, however, also 
contains a useful explanation of the notion of a strategy that is consistent with mine below: ``a strategy refers to a sequence of actions in a context where alternative sequences of (the same)  actions are possible''.\footnote{%
The restriction to the same progression of actions (see~\cite{BergstraP2009} where the notion of a progression is preferred over the notion of a sequence in the setting of actions)  was inserted by the present author. 
Moreover, I would prefer the following rephrasing of this statement: 
``a strategy refers to a premeditated progression of actions in a context where
alternative premeditated progressions of the same actions are possible''.}

 In~\cite{Shyrokov2010} multi-threading is studied against the background of another task (driving a car in a simulator). Multi-threading is found in dialoges  between two persons where different threads concern different themes. In particular threads may interrupt other threads allowing interrupted threads to be resumed in due time. 
 
 Below I will not allow that threads are interrupted by other threads.\footnote{%
 In multi-tasking, however, a task may interrupt another task from the same multi-task, and the interrupting 
 task may subsequently cause the resumption of the interrupted task. 
 More generally interrupts are possible between  concurrent processes (see~\cite{BaetenBK1986}).}
 External interrupts may occur, and such interrupts may  trigger a so-called contemplating phase of the interrupted thread (say $t$) after with a switch to another thread (perhaps an interrupt handler) may take place. 
If resumption of $t$ is envisaged a pseudo-switch to $t$ followed by a pseudo-switchback (pseudo-resumption) 
may be considered.\footnote{%
 An example of interrupts is as follows: agent $A$ is walking (task $w$) and talking (task $t$) at the same time. This is an instance of multi-tasking. If $A$ stumbles and as a consequence temporarily stops talking that constitutes an interrupt from $w$ to $t$, whereas if $A$ enters a complicated passage in talking $A$ may halt walking in order to be more focused on talking. The latter event constitutes an interrupt from $t$ to $r$. In both cases the interrupted task may be resumed after some time. In this example it is less plausible to view resumption of either task as the consequence of an action (resumption command) in the other task which suggests that modelling the situation with a third task with an executive role and viewing the issuing of resumption commands as part of the role of the executive task may be plausible. It should be noticed that an alternation of walking and talking (for the same agent $A$) may also occur in a context of multi-threading, for instance if $A$ walks towards $B$ with the intention of speaking to $B$ subsequently performs some talking to $B$ and then walks away. Both episodes of walking may be considered part of a thread ``physical motion'', and planning may precede strategic interleaving of that thread with a thread ``oral communication''.}

In~\cite{SuhonenRS2012} multi-threading is merely a description of a design in which several paths are an option. 
Then actions shared by many paths will be performed with high probability. This mechanism is quite distant from my understanding of multi-threading. I will propose how to to use threads as used in computing, and in particular as conceptualized in the thread algebra of~\cite{BergstraL2002,BergstraM2007} as a metaphor for how a human agent may try to view itself as a multi-threaded system.

Unfortunately multi-threading in computer science is hardly more well-defined than in social sciences. Multi-threading is the preferred term for explicit concurrency in most program notations that offer such features. 
When being put into effect 
(I refer to~\cite{Bergstra2011c} for an attempt at defining that phrase) in a single core machine the interpretation is a matter of strategic interleaving, when put into effect on a multi-core system with each thread being processed by its own core, the scene becomes closer to multi-tasking. Multi-threading viewed as a computer programming feature seems to be intrinsically ambiguous because what actually (and predictably) happens when a program is effectuated may differ from what the programmer had in mind, or might have had in mind. Programmer's tolerance for this sort of difference (and implied ambiguity) is far greater in the setting of multi-threading than in the more classical setting of sequential (procedural) programming.

\subsection{Designing a story on PMTh}
Everyone who combines a professional function in a day time job while enjoying a separate private life must know
 how to deal with at least two threads. If someone is also engaged in a sports activity  involving specific skills and experiences a third thread may be present which aggregates the actions relating to that sport. 
 
Both professional life and private life may further be split in different threads and so on in a hierarchical manner. 
I conclude from these examples that multi-threading is a nowadays a common feature for many individuals, perhaps 
more often than not in an unconscious manner, however. 

In this paper I intend to describe the role of multi-threading for a human agent and to 
specify PMTh in a terminology taken from (in part my own) work in informatics.

Rather than a theory on PMTh, with predictions and tailor made tools for design, analysis and maintenance,  
I will propose a ``story on PMTh'' from which a theory might evolve in due time. The story may provide a reader with
concepts and cognitive tools regarding PMTh which are helpful for dealing with their own or other persons' PMTh in 
more explicit, and perhaps more productive, ways.

\subsection{A thread algebra model of threads and multi-threading}
\label{TAproperties}
Thread algebra provides a theory of threads and multi-threads that I propose to use as the basis for the story on PMTh.
The thread algebra model of threads and multi-threading may be informally characterized by the following claims and observations.
\begin{enumerate}
\item Individual threads are extracted from instruction sequences. The latter may be understood as manifestations
of  plans. Thus at the base level threads represent sequential processes resulting from the effectuation of sequential plans. This idea has been put forward in~\cite{BergstraL2002,BergstraP2009} and it was applied in~\cite{BergstraM2012} and in~\cite{BergstraM2014}.

\item The interaction between thread and an environment is given by two operations: use and apply. A thread uses 
services in order to create a behaviour that is applied to its operating environment in order to achieve certain objectives. This approach to separation of concerns, taken from~\cite{BergstraM2007}, allows one to distinguish how a thread works from what it is meant to accomplish.
\item A multithread (also called thread vector) 
is a concurrent composition of threads, it results from applying a so-called strategic interleaving operator to a  thread vector (I refer to~\cite{BergstraM2007} for strategic interleaving).
\item 
It is plausible to have a hierarchical structure of multi-threads. (See~\cite{BergstraM2006}).
\item 
The use of strategic interleaving rather than arbitrary interleaving is characteristic of multi-threading, although in a particular case the strategic interleaving operator may allow run-time variations, which one may specify or describe by working with a family of strategic interleaving operators rather than only a single one.
This view has been developed in~\cite{BergstraM2007}.\footnote{%
In~\cite{BergstraM2007} the so-called cyclic strategic interleaving plays a key role, mainly because it embodies the 
most straightforward idea  on how to design an interleaving strategy. Remarkably, in~\cite{SalvucciT2008} which 
aims at a description of human psychology cyclic interleaving is given a central role as well.}
\item
While in the formalized setting of description of programming language semantics (see~\cite{BergstraM2007}) strategic interleaving is specified by means of strategic interleaving operators, in a less formal setting I prefer to speak of strategic interleaving 
policies (SIP)  rather than  of strategic interleaving operators. This view of concurrency may be referred to as 
SIP driven multi-threading. With SIP driven it is expressed that a policy rather than accidental race conditions determine which threads will be active, for how long, and in what order.
\item 
Choosing a strategic interleaving operator  from a pool of operators  as well as ways of transferring control from one thread to another thread according to a given strategic interleaving operator is delegated to a strategic interleaving policy. SIP driven concurrency of multi-threads is an attractive intuition that we may transfer from computer programming into other fields of application. 
\item 
Multi-threading is especially suited for cases where threads show alternating activity, excluding one another in time,  and where more or less sophisticated choice mechanisms, 
sometimes depending on the evaluation of complex conditions, may be employed to determine when to carry on with another thread. This view of a multi-thread differs from the intuition of multi-tasking where different tasks are physically performed simultaneously.\footnote{%
Car driving and simultaneously handling one's smartphone is a notorious (and problematic) example of multi-tasking 
in the case of human agents.}
\item 
If the agent performing a multi-thread is a human agent I will speak of personal multi-threading (PMTh);
if a human agent performs multi-tasking I will speak of personal multi-tasking (PMTa). It is reasonable to view 
a multitask as a single thread in a multithread.
\item
If one prefers to model a human agent as a multi-processor thread mobility and migration comes into play. 
Such mechanisms can be easily formalized and explained in the setting of thread algebra
(see~\cite{BergstraM2007}). In this case a distributed strategic interleaving must be used.\footnote{%
If a multiprocessor system is used in such a way that thread vectors are partitioned 
into groups (sub-vectors)
and each sub-vector is effectuated in its own timing by one of the processors the interleaving policy should
be classified as an arbitrary interleaving rather as than a strategic interleaving.}
\item If one prefers an  object oriented style of modelling a single agent thread algebra can be 
used as well. A probabilistic version of thread algebra can easily be defined.
\end{enumerate}

\subsubsection{Multi-threading as a conceptual tool}
I suggest to use multi-threading as a conceptual tool. Although the equations of thread algebra  and
its  variations constitute the foundations of a rigorous bookkeeping of thread dynamics and no more than that,
the shape of thread algebra can also be understood in  more informal ways. 

This informal understanding of a formalized system (for threads, multi-threads and strategic interleaving) 
may be compared to one's grasp of formal logic where a proof 
system may explain logical implication so convincingly that one may 
feel almost compelled to apply that attractive notion in informal 
daily practice simply ignoring the fact that it almost never applies. 
Logical thinking can be used as a positive label of informal practice. But the question remains 
which logic is meant.

In practice reasoning patterns  are probabilistic, plausibility reigns over certainty,  
probabilities are often unknown, reasoning is hampered by defaults and priorities, 
and  inconsistency is the rule rather than the exception.\footnote{%
The connection between implication and causality is often unclear. 
Practical implications are mostly asserted and only in rare cases implications 
are formally derived from other information at hand.}
Upon becoming aware of all of these complications  one may feel invited to design one's own ``philosophical'' logic taking all of these matters simultaneously  into account, thus rendering the original proof system into an almost naive status, comparable to the admittedly naive status of the strategic interleaving policies that have been put forward in~\cite{BergstraM2007,BergstraM2006,BergstraM2008}. 

\subsubsection{Virtues of hierarchical  multi-threading}
With a focus on an informal approach to agent behaviour the notion of hierarchical multi-threading 
(HMT) is attractive for the following reasons.
\begin{enumerate}
\item 
Strategic interleaving is an intuition. One may think of applying a strategy even while in fact that is not true.
The idea of possessing (performing) a well-managed hierarchically structured family of threads is attractive.
\item 
The plans that are effectuated by threads can be extended (and modified) as a part of thread management. This mechanism is outside the scope of thread algebra but the intuition is clear. I suggest condensation as a metaphor, 
with the thread as the liquid phase of a vaporous medium of goals and objectives. Effectuation of the a thread is comparable to freezing its liquid phase. 
Condensation progressively  transforms an unstructured set of goals and objectives into
a rigid instruction sequence. The instruction sequence in turn is effectuated thus producing a progression of actions which 
plays the role of a frozen plan.
\item 
Strategic interleaving places the intuition of mutual exclusion of thread activity upfront. It focuses therefore on the conflicts that must be resolved whenever the meta-step of switching to another thread is ahead. In addition, however,
the concept of strategic interleaving (and in particular its probabilistic versions) permits a gradual shift towards the intuition of truly concurrent multi-tasking that is much less sensitive to the choice of an interleaving strategy up to the point that arbitrary interleaving becomes a preferable model.
\item HMT with strategic interleaving allows for many decorations of its components with additional information. First of all instead of a single strategic interleaving one may use a family of strategies and a policy for selecting an appropriate member of that family when needed. Secondly the individual threads may all be equipped with private state spaces. These state spaces may maintain formal and quantified attributes as well as informal ones. The managing thread sees to it that effectuation in all other threads is smooth.
\item 
The arbitrary interleaving policy on which process algebras are based (see for instance 
muCRL2 of~\cite{GrooteM2014}) fails to accommodate these informal 
connotations mentioned above which strategic interleaving policy  supports rather well.
Arbitrary interleaving has a bias towards the description of concurrent multi-tasking and it requires an addition of the
strategic aspect to the parallel composition operators. While for computing arguably concurrent multi-tasking is most prominent and arbitrary interleaving is prior to strategic interleaving that seems to be different for conscious human activity.\footnote{%
In~\cite{BergstraM2007} it is argued that for understanding the concurrency in programmed systems with parallel features
the intuitions of strategic interleaving is more convincing than the intuition of arbitrary interleaving.}

\item Thread algebra presents multi-threading in the form of a theory that is partial, incomplete, and unfinished, and which incorporates almost indefensible simplifications (fixed interleaving policies, 
threads emerging from instruction sequence effectuation)  in order to produce a mental picture
of a system that is easy to grasp. These deficiencies render thread algebra unattractive for theoretical computer science where one is used to more fundamentalistic and more elegant formalisms. 
At the same time these deficiencies seem to be helpful for visualising or imagining the intuitions at hand (at least in case 
of human agents).

\item It is plausible that a thread is never fully inactive and that some background mental processing of it 
takes place even if another thread is active. This may be needed for a human agent to stay aware of the thread's existence and to maintain motivation, goals, objectives and relative priority w.r.t. other threads in the thread vector.
\end{enumerate}

\subsection{PMTh versus PMTa}
Making a sharp distinction between PMTh and PMTa may not be possible in al cases. In the case of PMTh at some stages thread effectuation may be similar to task effectuation and a gradual transition of PMTa may be in order. 
This may happen for instance if one is writing a document in the context of thread $t$ and needs to participate  in a telco regarding an activity in thread $r$. Then it may be practical to keep writing and to handle the phone call simultaneously by way of multi-tasking. This aspect of the present paper is perhaps underdeveloped. If thread agency involves tasks amenable for multi-tasking it is clearly less plausible to focus exclusively 
on multi-threading as a modelling method. 

\subsubsection{Sessions as a distinctive feature}
Personal multi-tasking (PMTa) is different from PMTh in that it provides room for the notion of a session. 
For instance during a telephone call an agent may alternate between speaking and listening, 
and in between an agent may interleave actions belonging to another task, say $t$. 
In such a case the agent  switches to $t$ and back while staying within the same telephone session. In PMTh there is no such notion of a session. Once a switch from say thread $r$ to thread $t$  to another thread has been made a subsequent switch back  to
$r$ entails a restart from scratch rather than a resumption of one or more sessions that have been interrupted in order
to switch to $t$.

A prescription that one should not write a tweet while car driving makes sense under quite specific conditions only:
\begin{enumerate}
\item  tweeting and car driving are both considered to be subsumed in the effectuation of different tasks (say message communication and physical transportation respectively),  
\item successive actions of producing and sending a tweet are grouped together in one session, 
\item successive actions of a car trip are aggregated in another session, 
\item both sessions can be referenced by name
or otherwise somehow implicitly, and
\item  it can be formulated as a requirement in advance that both sessions should not overlap in time.
\end{enumerate} If both tasks are considered threads simultaneous effectuation is an impossibility 
which need not be forbidden for that very reason. Aggregating actions from different activations of the same threads
into a session is implausible because switching away from a thread brings all activity of that thread to a halt.

\subsubsection{Contemplating activity}
Below I will speak of contemplating activity if when active within one thread an agent contemplates the feasibility of switching to and being active in another thread. Similarly one may think of external task activity if when active in a thread an agent simultaneously performs a task which logically might be best understood as belonging to another thread. Strictly limiting the focus to multi-threading implies that external task activity to some extent can be counted
as belonging to a thread proper.

\section{Multi-tracing: threads ex-ante}
An agent looking back to an extended history may wish to distinguish threads of activity that together constitute, 
and explain, that history. Such threads are ex-post threads. Events separated by many years and causally unrelated may be collected in a single (ex-post) thread. 

For instance if one wins a lottery three times these events may be grouped together in a thread 
named ``good luck''
 in spite of the 
absence of any causal relation between these events and 
even in the absence of any planning involving more than one lottery participations at a time. Multi-threading and strategic interleaving concern ex-ante threads, 
threads which are at any stage at most in part history and which constitute at least to some (non-trivial) extent  plans about the future, or prototypes thereof.

Ex-post multi-threading concerns the reverse engineering of an agent's history into a collection of independent subhistories. No mechanism for choosing between actions and threads plays a role, such mechanism can be revealed by historical investigation only. Ex-post threads may alternatively be called revealed threads. 

Different observers may describe the history of the same agent with different ex-post multi-thread. In an ex-post multi-threads threads that have come to an end may play an important role because for such threads an assessment can be made about the extent to which initial goals have been achieved or even exceeded.

In concurrency theory one often uses the notion of a trace for an ex-post thread, or rather for the interleaved past of all ex-post threads of an agent or a system. I propose to use the term multi-tracing as an alternative term for ex-post multi-threading, and to have ex-ante multi-threading as the default meaning of multi-threading.

\subsection{Multi-tracing and multi-threading as competing views}
An agents who looks back at its own history as a multi-trace of success stories may be happy about that interpretation of its personal history and at the same time the agent may need to forget or ignore this optimistic
view altogether (or temporarily) in order to focus on what lies ahead. Multi-threading may overestimate future problems and may undervalue past achievements, multi-tracing may have the opposite effect. Through a multi-tracing perspective an agent may determine important statistics of its own ways of working. For instance if on average a trace  in the 
agent's past needed 10 years from initiation to a successful completion, it may be unrealistic to expect threads of a multi-thread to produce results much faster.

Multi-tracing and multi-threading provide a matrix view of an agent, where multi-threading concerns ``how to'' and ``what next'' (thus imposing a prospective bias) while multi-tracing redraws the picture in preparation of assessment and evaluation (thus imposing a retrospective bias). One may imagine an agent oscillating between these views, or of one accepts priority for multi-threading over multi-tracing one may simply introduce a thread the objective of which is to develop, update, and maintain a useful multi-trace for describing (and rationalising) an agent's personal history.

\subsection{Progressions: between traces and threads}
A multi-thread picture of an agent's future features branching just as a multi-trace picture of an agent's past. Complementary to what will happen as a selection from what might happen as codified in multi-threading, an agent 
may encode its view on what has happened as a selection from what might have happened in a multi-trace picture of its past.

There is no sharp or even trichotomy between past, present, and future, however, in the language of traces and threads. Following~\cite{BergstraP2009} I will speak of a progression (of actions and/or events) if there is
a sequential order, that is a linear order that supposedly corresponds with temporal order, and if there is no definite commitment to past or future (initial steps in a progression may belong to the past, later steps may still lie ahead, and for intermediate ones that may be open). A progression abstracts from  branching for instance by choosing a particular history (trace) and committing to a specific plan (a progression through the branching structure of future actions and events).

\subsection{Multi-tasking ex-post}
For multi-tasking just as for multi-threading past and future can be distinguished. If an agent is fined because of simultaneously being engaged in a session of driving and in a session of social media use, it is unavoidable to think (and speak) in terms of past tasks and past sessions. Multi-tracing is as good a tool for ex-post multi-tasking as it is for ex-post multi-threading. A consistent language about these issues results if one assumes that by default
multi-tracing represents ex-post multi-threading while only when explicitly indicated it may 
be used in the context of ex-post multi-tasking.

\subsection{Parametrized concept development}
Previous results of work to which I have contributed are imported in the present work on personal multi-threading thus positioning it in the line of a multi-trace of previous work. As these traces have developed somehow in an accidental manner alternative outcomes might have produced alternative building blocks of equal or even greater value. I hold the view that my story on PMth is parametrized by
several units of existing work which have been used (that is substituted as an actual parameter) while alternatives,
including alternative solutions for similar problems found in the literature, might have been used as well.

An important dichotomy is between endurants and perdurants. In the context of this work agents are considered 
endurant entities, which means that an agent can be observed as a whole. Threads and processes however are perdurants in the sense that at any moment of time one only observes a snapshot of a thread or a process.\footnote{%
In a process calculus and in a process algebra, and in a thread calculus as a well as in a thread algebra processes or threads are perdurant entities while process expressions or thread expressions that can be used to describe threads and processes respectively are endurant entities. It is rather difficult to maintain a clear type distinction between endurance and perdurants in a process theory or in a theory of threads.} 
Using the contrasting pair of endurance and perdurance, constitutes an
import form philosophical ontology for which alternatives may be found. The precise form of ontology imported in a story
about PMTh may be considered an actual parameter that has been chosen from a range of alternatives while the story itself
viewed at a higher level of abstraction features a formal parameter for a piece of ontological content.

Besides a philosophical distinction between endurants and perdurants other imports from nearby theories
in PMTh have occurred. My story on PMTh has several other constituents that also may be viewed as actual parameters that have been used to instantiate formal parameters of the story on PMTh and for which alternative actual parameters might have been used alternatively.

The view on PMTh as put forward below uses several developments imported from other work. More specifically:
\begin{enumerate}
\item Thread algebra with strategic interleaving is used for the conceptualization of multi-threading.
\item Outcome oriented decision taking (OODT from~\cite{Bergstra2011a}) is used to conceptualize the notion of a decision.
\item Program algebra and instruction sequences (\cite{BergstraL2002}) constitute my favourite tool for conceptualising the notion of 
a deterministic plan, while  a thread is considered an appropriate description of a plan that is being effectuated 
(thread ``is'' PuE: Plan under Effectuation.).

\item Stratified sourcing theory  (see~\cite{BDV2011c}) is used as a basis the notions of source, sourcement, insourcing, outsourcing.
and backsourcing, and 
\item  The notion of conjectural ability from~\cite{BDV2011b})) is used to explain why and when the given view on PMTh might be profitable.
\end{enumerate}

Each of these imported viewpoints or theories concerning specific themes that I consider to be of relevance to PMTh 
may be replaced by alternative and somewhat different theories. As the notion of PMTh  depends only
marginally on a particular choice in any of these parameter theories on may feel disinclined to fix any of these 
parameters. At the same time I prefer to work in a context where such choices have been made because it provides a useful platform of definitions for ingredients that are in need of a definition before being 
used as building blocks for a description of PMTh.

Of course each of these choices for a parameter theory also constitutes a risk. If that choice is considered unconvincing, either in general or in this specific context, that judgement will unavoidably reflect negatively on the proposed conceptualization of PMTh.

\section{Hierarchical thread vectors: an architectural model}
In this section I will describe how to use HTVA (hierarchical thread vector architecture) as a tool for describing an agent's control state as far as multithreading is concerned.

\subsection{What makes a single thread}
A single thread incorporates activities that have something important in common. Commonality may range from goals, mission, and strategy to workflow uniformity, or may have to do with the context of interaction with other agents. Not much can be said in an a priori manner for agent $A$ it may be obvious that a job based professional life can be captured with one or more threads only consisting of job related activities, while for another agent, say $B$, a single thread may group together activities that $A$ would consider private with activities that $A$ would consider job related.

 Trivially one may subsume all activities of agent $A$ in a single thread. That leads to the trivial HTVA for $A$. Different observers of $A$ (including $A$ itself)  may have different views on how to decompose $A$'s behaviour in different threads. 
The stability of a thread during a phase of an agent's existence is connected with the stability of other threads. For most agents $A$, each non-trivial HTVA for $A$'s activities is arbitrary in the sense that it represents a design by that has ben constructed part in terms of reverse engineering. If a particular non-trivial HTVA is considered unconvincing,
it may be replaced by the unique trivial HTVA, in preparation of its redesign.

In~\cite{Bergstra2012c} I have used the notion of a single thread as a high abstraction for the specification of the process of selling a valuable item. For that paper the distinction between thread and task is immaterial, however.

\subsubsection{Thread vectors}
A thread vector is a finite sequence of threads. A thread vector is a hierarchical thread vector of depth one.
A hierarchical thread vector of depth $n+1$ is a sequence the elements of which are either threads or 
hierarchical thread vectors of  depth $n$ or lower. The latter elements are called subvectors of the thread.
Threads that are collected in a subvector are supposed to have important aspects in common. If a HTVA shows 
hierarchical threads I will speak of a multi-level HTVA.

In the thread algebra model
the thread vector is such that the head of the list (vector) contains the thread which is now active, if any. Switching
to another thread involves a permutation, sometimes merely a rotation, within the thread vector.
Threads may have an independent numbering or possibly mnemonic naming  

A hierarchical  thread vector may be viewed as a very simple architecture of the state of an agent, with each thread comprising a specification of a part of the agent's future behaviour.\footnote{%
Instead of future behaviour one may speak of an agent's personal dynamics.}

Modelling a person's behaviour with multi-threading does not imply an assumption that 
human beings observably (or internally in a nonobservable manner) operate in a multi-threaded fashion, 
and more particularly it does not presume any assumption that the
thread algebra model of multi-threading is particularly suitable to collect and describe empirical facts
about human multi-threading assuming provided that agrees that such a thing exists to begin with. 

Of course the possibility of conceptualising human behaviour in terms of multi-threading is by no means new, but focus seems to have been on multi-tasking rather than on multi-threading.  Below I will contemplate how possible applications of this architectural model may be found. Primarily it is a matter of developing a language for conceptualising a 
certain range of phenomena rather than to develop models allowing predictions amenable for empirical confirmation.

\subsubsection{HTVA design: descriptive, prescriptive, or suggestive}
Modeling an agent's behaviour by means of a HTVA is probably best viewed as a stepwise process. 
The process depends on the objectives of developing a HTVA. For PMT the case that agent $A$ designs its own HTVA is of particular importance. Dissatisfaction of $A$ with its current  HTVA, say $a$ can arise 
in several ways. Threads may be felt as incoherent, demotivating, or as too much loaded with problems, and many other psychological factors may come into play. 

I hold that supporting an agent with the stepwise design though subsequent
regions of it own HTVA constitutes a service for which an agent (say $B$) may develop specific abilities.
Awareness of the story of PMTh is supposed to provide empowerment of $B$ as a PMTh consultant by way of
providing certain specific conjectural abilities (see~\cite{BDV2011b} for that notion). Working out the details of the package of conjectural abilities that is supposed to come with the story of PMTh is left for future work.

A PMTh based architecture of an agent's dynamics (that is a HTVA)
must not be understood as descriptive in a scientific manner, 
the model may be wrong. It must not be considered as prescriptive either, there is no reason why an agent ought to be organized in that manner. 

Instead a HTVA for an agent may be considered a conceptual tool, or in somewhat less friendly language, 
it may be labeled as suggestive. It proposes a view that may be useful, but no guarantees to that extent are provided.
What makes a useful HTVA is a subject for further work. Some remarks are in order, however.
\begin{enumerate}
\item A crosscutting HTVA design combines different activities that may be at first sight classified quite differently in single threads. For instance a thread ``hobby'' may include part of one's work, while one of the threads constituting $A$'s
professional activities may include aspects that would plausibly be classified as private. Crosscutting HTVA designs may be useful if more conservative designs, like\\
 $<<$private$>$+$<$hobby$>>$+$<$work$>$+$<<$family$>$+$<$friends$>>$\\
don't produce helpful or stable results.
\item $A$'s HTVA must preferably be helpful in strengthening its operations. Well-named threads can 
reinforce a sense of identity. Negative emotions and experiences may be concentrated in certain threads in order not
to weaken the operation of other threads.
\item Threads named in $A$'s HTVA must invite $A$'s activity and must preferably 
provide $A$ with a profile that can easily be communicated and which allows giving a rationale for $A$'s activities
in spite of a chaotic and unstructured impression which these activities may make when superficially inspected.
\item $A$ must be able to stop and later resume activity for each of its threads. However important a thread may be for $A$'s well-being or even survival, interrupts may occur and the competence to pay attention to other threads needs to be assessed and found in good order.
\item Ideally an agent feels at home int its HTVA, like a person might be happy with the distribution of functionalities 
and corresponding furniture over the different rooms of a private home. Here the rooms may be compared to the threads of a person's HTVA, and an agent's feeling at home might correspond to the ability to be at ease in different rooms for prolonged periods in combination with the absence of a permanent urge to move furniture around and to change the functionality of rooms.
\item Each thread also defines a perspective from which $A$ perceives its existence. Thus when active in thread $t$, $A$ has a view on thread $r$ which may change after $A$ made a switch to thread $s$. Preferably there is no overall perspective that governs the perspectives per thread. Each thread must accommodate a complete philosophy regrind the entire existence of the agent at hand. Each important viewpoint concerning the agents existence should be found a home in one of the threads of its HTVA.
\end{enumerate}

PMTh and HTVA design may be viewed as a proposal in the area of ergonomics. In~\cite{Charytonowicz2009} 
a historic 
survey of ergonomics is given which is compatible with the idea that ergonomics reaches out beyond physical aspects of the workplace and of physical and mechanical circumstances in private life to encompass well-being oriented
optimization at a psychological level. Under such a definition PMTh may be viewed as an attempt to contribute to ergonomics.

\subsection{An executive thread}
One thread may be assigned the task to take care of the SIP including HTVA maintenance. This thread may be called
the SIP executive thread or briefly the executive thread. In particular creating 
new threads and removing threads whose existence has become a disadvantage is work for the executive thread. 
The executive thread also hosts the
mechanisms for effectuating a particular SIP. Choosing which strategic interleaving to use for some 
period of time and managing the transfer of control between threads is work for the SIP thread.

In the absence of an executive thread all threads must each perform control related tasks in an alternating manner.
I prefer to consider an HTVA without a dedicated executive thread and to distribute thread vector management of 
all threads. A similar point of view can be found in~\cite{SalvucciT2008}. 

\subsection{Thread attributes}
Several attributes may be needed to obtain a useful level of 
expressiveness for a HTVA. An obvious attribute is the activity flag.

A thread may be active (flag on) or passive (flag off) or in between (contemplating activity). The lower its degree of activity the more it is needed that the dynamic state of the thread is stored (saved) in the state of a HTVA support system 
in order to allow resumption of the thread after a thread switch.
\begin{description}
\item[thread name.] The name of a thread allows making reference to it when reflecting about it. 
In stepwise HTVA design thread names may be replaced by names that are expected to provide a better fit with 
the aggregations of actions and tasks embodied by the various threads.
\item[state space.] The state space of a thread comprises the totality of all objects the state or existence of which is
which affected by actions of the thread. Ideally the state space is constant during the existence of a thread. If the thread's actions have a deterministic effect on the states in the mentioned state space and actions of different threads do not interfere the state space may be split from the thread and a single state space may be acted upon by a multi-thread. 

In on thinks in terms of use and apply the state space for a thread my be decomposed into an internal state which
is used by the thread and an external state to which the thread is applied.
\item[state.] The state is at each time a member of the thread's state space. It may be the case that some subsets of the thread vector have a shared state space each.
\item[mission.] (Qualitative description of) ``what the thread is good for'' from the perspective of the agent.
\item[targets.] (Description of the) ``results that thread effectuation is supposed to deliver''.
\item[objective prominence.] How important is the thread for the agent, judged from outside.
Measured on  a 1,..,5 scale, the average over all threads must at all times be 3.
\item[subjective prominence.] How important is the thread for the agent viewed from inside. 
Measured on  a 1,..,5 scale, the average over all threads must at all times be 3.
\item[workload fraction.] The sum of workload fractions of threads must be 100\%. 
The following ramification may be needed:
\begin{itemize}
\item subjective workload fraction,
\item objective workload fraction,
\item intended workload fraction,
\item expected (from outside) workload fraction.
\end{itemize}
\item[effectiveness.]
Between these quantities there are no required quantitive connections, but anytime each must count to 100\% 
when counted over all threads in existence.

Similarly, effectiveness of a thread, i.e. the degree of its success in reaching stated objectives may vary 
through time. There is no summing up to 100\% in this case.
\begin{itemize}
\item subjective effectiveness,
\item objective effectiveness,
\item intended effectiveness.
\item expected (from outside) effectiveness.
\end{itemize}

\item[flow.] Flow is a sense of steps/actions naturally following one another. A positive experience of flow stabilizes thread operation.
\item[satisfaction.] Besides flow the feeling of satisfaction that thread agency may create promotes continued thread activation.
\item[identification.] An agent may derive a sense of identity from its thread effectuation activity.
\item[clarity of ``what next''.] In downhill skiing it may be so obvious where to go next that this very clarity 
creates a sense of flow that constitutes a
significant factor keeping the thread active. In finding one's way through a labyrinth it may be the other way around and a lack of clarity on what next may create a sense of standstill.

\item[other attributes.] The following attributes each have  internal, external, intended, and expected versions:
external visibility,
rewardingness,
appreciation,
rationality,
avoidability,
removal risk,
good order,
preciousness,
subthread structure.
\end{description}

\section{Thread switching}
When one thread is deactivated and another thread is activated a switch takes place. 
If present the SIP thread indicates when and how thread switching takes place. 
 Given this intuition, however, 
one may investigate the
variety of mechanisms that may come into play just before, during, and after a switch. 
I will assume that $A$ is the (name of the) agent whose behaviour is supposed to be captured 
by the multi-thread under consideration.

A switch can be towards contemplation (contemplating activity) or towards (real) activity. 
I will write C-switch and A-switch.
It seems plausible only to allow A-switches from contemplating activity to real activity of the same thread. It
is also plausible to have C-switches from a thread $t$ to another thread $r$ only if $t$ is in contemplating mode of activity.

\subsection{Thread switching as a meta-action}
Because we assume that actions take place during thread effectuation, a switch is not an action.
A switch is considered a meta-action, that is a change in the pattern of action that may be observed 
by an external observer but which in itself will not qualify as an action.

As a rule I will assume that a thread switch is a meta-action that takes place on behalf of the agent.
It may involve choice (the agent deliberates about which thread to switch to and makes a 
conscious choice on that matter), or action determination (the agent turns out to switch to another thread
as a consequence of the real time flow of events but without preparatory conscious deliberation).

\subsubsection{Thread switching caused by (direct) decision taking}
I will not say that a switch performed by agent $A$ indicates that $A$ has taken the decision to engage in 
a thread switch. Following the terminology of OODT (outcome oriented decision taking, 
see~\cite{Bergstra2011a,Bergstra2012a,Bergstra2012b})
decision taking requires a protocol which terminates with the production of a decision outcome. Putting the decision outcome into effect must cause the switch if the switch is to be understood as resulting from decision taking.

If $A$ is a multi-agent instead of a single agent it is plausible that a switch results from internal decision taking. Under the assumption that $A$ is a single agent, which is always valid in the case of PMTh, 
internal decision taking is rather implausible as a cause for thread switching.
Single agent thread switching is more likely to be a part of decision taking-less activity (DLTA). Choice rather than decision is the most plausible cause of a thread switch. In the case of multi-tasking, however, besides choice als action determination (see~\cite{Bergstra2012b}) plays a role as a cause for task-switches.\footnote{%
In~\cite{Bergstra2012b} the possibilities of internal decision taking for a human agent are considered from the perspective of OODT with the conclusion that this mechanism cannot be excluded on theoretical grounds.}

Summarizing it follows that thread switching caused by direct 
decision taking is implausible in the context of PMTh. In other words, decision taking-less agency
is a plausible mechanism for thread switching.

\subsubsection{Thread switching caused by indirect decision taking}
However, the agent $A$ may behave in such a way that an agent $B$ external to $A$ proceeds 
with taking a decision that induces changes in the environment of $A$ which then lead $A$ to engage in a 
 switch to say thread $t$. 
 
 This may happen for instance if $A$ asks $B$ for permission to carry on with 
thread $t$ and if $B$ shapes its response in the form of a decision to give a the permission of which $A$ is in scope. 

It may also be the case that progress on thread $t$ is blocked for $A$, and that 
$B$ takes a decision $d$ and by putting the decision outcome $O_d$ into effect one or more agents
(not including $A$) in scope of $d$ produce a new state in which $A$ can progress with $t$ because the blockade as ceased.

I will refer to the the latter mechanisms  as thread switching caused by indirect decision taking.

\subsection{Why thread switching may be hard}
Thread switching from say thread $t$ to thread $r$ may be difficult to perform for an agent $A$ for a plurality of reasons:
\begin{enumerate}
\item 
$A$ may have become used to running $t$ (satisfying experience of flow) 
and $A$ may feel very uneasy about phases where no 
progress on $t$ is made (fear of forthcoming lack of flow). One may think in terms of $A$'s attachment to $t$, 
the attachment being unsafe if $A$ is unable or unwilling to leave $t$ (that is to switch away from $t$) in order
to perform $r$ even if effectuation of $r$ is temporarily significantly more important to $A$.
\item $A$ may lack the motivation to begin acting upon $r$.
\item $A$ may not know how to get started with acting on $r$. (Lacking clarity on ``what next''.)
\item $A$ may not be sure that the state of $t$ can be saved in such a manner that resuming the 
thread on a future occasion is practical.
\item $A$ may be worried that after leaving $t$ it may not return to it and $t$'s mission will remain unfinished 
in spite of extensive investments of time and other resources in it.
\end{enumerate}

Remarkably setting up and maintaining a HTVA in order to self-apply  PMTh is likely to be most useful for 
those agents who find efficient thread switching most difficult. Some agents may have a natural way of PMTh in no need of being made explicit and being reflected upon.

\subsubsection{A thread separation paradox}
If $A$ contemplates a switch from thread $t$ to thread $r$ and finds (makes the assessment) 
 that its preparations concerning $r$ are still insufficient
to allow that switch, there seems to be no other option than to improve 
these preparations before making the switch to $r$. 

This seems to imply that detecting the need for these preparatory steps such preparations can be done 
while  thread $r$ is still inactive. That is implausible, indeed almost paradoxical, 
as it implies that each thread has the capability of analyzing possible 
progress from any state for all other threads. I will speak of the thread separation paradox to express that in a flat
thread vector architecture in which each thread is either active or inactive at any moment a switching mechanism either cannot impose any conditions before switching to some thread $t$ or presumes that the capability to express and check such conditions is embodied in all other threads $r$ from which a switch to $t$ may occur.

\subsubsection{Contemplating activity phase and proper activity phase}
As a solution   for the mentioned thread separation paradox paradox I  propose to distinguish two forms of 
activity: ``proper activity'' and ``contemplating activity''.\footnote{%
In~\ref{TPS} below I will in addition distinguish ``pseudo-switched activity''.}
Thus rather than having a degree of activity between on and off there is contemplating activity 
as a light weight form of activity. A switch from $t$ to contemplating activity of $r$ is easily reversible and it
does not involve an exit from the flow of $t$ and it is hardly touched by other factors that make thread switch hard.

While proper activity of threads is mutually exclusive in time different and contemplating activity is 
mutually exclusive, two different threads can have simultaneous contemplating and proper activity. 
A thread switch from $t$ to $r$ may involve several consecutive phases of 
contemplating activity of threads different from $t$ before thread $t$ becomes properly inactive and  $r$ becomes properly active.

Once contemplating activity has been introduced as a modality of thread activity the possibility arises 
that contemplating activity becomes dominant in the sense that most turns allocated to threads are of a contemplating form rather than constituting of proper activity for that thread.

I will assume that a strategic interleaving policy only specifies switches to a proper activity phase, such switches are called proper switches, and that
in a proper switch from $t$ to $r$, a properly active phase of $t$ is ended just before the priorly active phase of $r$ begins.

\subsubsection{Thread switching readiness}
Af $A$ switches to thread $t$ it should be part of the contemplating phases that $A$ prepares itself for a subsequent switch to another thread. Part of each contemplation is reconfirming awareness of the entire thread vector. 
If a thread tends to become forgotten that fact must be detected. In its simplest form this requires at least two contemplation phases in advance of a switch from $t$ to $r$, contemplation of $r$ in order to check that proceeding
with $r$ is enabled, and contemplation of some other thread $s$ in order to see that a subsequent switch 
from $r$ to $s$ is plausible.

These meta-actions allow an on-the fly construction or adaptation of a strategic interleaving policy, 
or an initial segment of it.

\subsubsection{Procrastination and atychifobia}
In spite of the fact that the actions specified by thread $t$ may be quite important for $A$,  
$A$ may wait excessively long before resuming activity within thread $t$. This phenomenon when occurring in human agents is termed procrastination. There is a significant literature on procrastination.

A cause for procrastination in certain conditions may be atychifobia. 
Agent $A$ suffers from atychifobia if it often experiences an exceptionally strong fear of failure. 
If one assumes that $A$'s 
atychifobia with respect to tasks included in $t$ causes delay of switching to $t$ and if one assumes that this 
sentiment is worsened because $A$ considers thread $t$ highly important, the counterintuitive conclusion is
that $A$'s procrastination of resuming $t$ may be alleviated by reducing rather than increasing 
$A$'s subjective importance (prominence)  assessment of $t$.

\subsection{Motives for thread switching}
Carrying out a proper switch can have many grounds and causes. 
The thread vector architecture is irrational for $A$
 if switches never occur in $A$'s observed or expected behaviour. 
 So the occurrence of switches is implicit in the jargon of threads and
 If no switches occur a single thread architecture is a preferable model of $A$'s activity.
 
Having more than one thread may be favorable if modularization of $A$'s activities in threads leads to a gain in 
structure for each thread which outweighs the cost of maintaining a thread vector and a strategic interleaving policy, 
as well as a policy for preparing proper thread switches through contemplating phases of various 
threads, each of which may either become the next properly active thread or the one thereafter.

At least the following motives for a proper switch may come into play when a switch from $t$ to $r$ is made by $A$.
\begin{enumerate}
\item Fairness:  in order to prevent that $r$ remains untouched too long. Apart from that the discontinuation of $t$ has no grounds in how that continuation proceeds.
\item $t$ has become blocked and its continuation is temporarily impossible. In this case 
progress with the effectuation of $t$ is blocked by some condition, and $A$ must wait until the blocking condition has ceased.
\item A change in the perception of priorities has made achieving progress on $r$ more important that progress on $t$
compared to recent history where the properly active phase for $t$ could be motivated on the basis of its perceived relative importance in comparison with other threads from the thread vector.
\item A has become tired from pursuing $t$, and for that reason its processing of $t$ has become less effective, and $A$ needs some change (that is working on $r$ which should require a different kind of work) while 
expecting that in a later stage $a$ will be more effective after returning to $t$ once more.
\end{enumerate}

\subsection{Thread pseudo-switching}
\label{TPS}
Unfortunately the view that concurrency in agent $A$ must take either one of two forms, multi-threading or multi-tasking, is too simple. In the multi-threading view a useful feature is yet missing. This feature allows the agent to carry on with some thread, say thread $t$, and then to choose to take a progression of actions $p$ (some logical task) from another thread, say $r$, and to perform $p$ as if it were a part of thread $t$. Thus $A$ will not make a full switch but 
rather apply a temporary interrupt of its effectuation of thread $t$ for doing $p$ while staying in the mindset of $t$.
I will refer to this mechanism as a pseudo-switch rather than a (full) switch. Each pseudo-switch must be followed by
a pseudo-switch in the opposite direction (a so-called pseudo-switchback or pseudo-resumption). 

Pseudo-switching may be relevant if $p$ (and perhaps other 
progressions from $r$) need to be effectuated with urgency while
$t$ still needs to be felt by $A$ as its top-priority. If subsequent progressions from $r$, say $p_1,p_2,..,p_k$, are
performed by $a$ after successive pseudo-switches (followed by switchback's)  from $t$ to $r$ then that succesion must not be considered a session of $r$. If it were to be considered a session either a full switch needs to be made from $t$ to $r$, or  instead of multi-threading a multi-tasking model is more appropriate to model the architecture of that part of $A$'s activities.

\subsubsection{Contemplating activity as an instance of pseudo-switching}
It seems plausible to view contemplating activity also as an instance of a  pseudo-switch. In some cases that is convincing, for instance if $p$ needs to begin with a planning phase. 
In other cases understanding contemplating activity 
in $t$ about $r$ is a less convincing, in particular if contemplation is vague and informal in comparison to planning activity within $r$ proper. There seems to be a continuous scale: 
on the one hand each thread $t$ is be attributed the capacity to 
allow reflection about the potential of a switch to any other thread $r$, while on the other hand detailed  thought experiments concerning the planning of progressions of actions of $r$ which are informed about all technicalities of the objectives that $r$ is aiming at can be performed best as a part of $r$, that is after a pseudo-switch.

\subsubsection{Motives for pseudo-switching}
The primary motive for a pseudo-switch (from $t$ to $r$) being that some actions or progressions of actions
 of $r$ are felt in need of urgent attention while running thread $t$, a secondary motive appears. $A$ may feel not at ease with the entirety of tasks located to thread $r$ and may for that reason be disinclined to switch to $r$. At the same time $A$ may feel comfortable with parts of the activity allocated to $r$ and may prefer to perform this tasks without taking on board the task of reflecting about a planning of the whole package of activities that $A$ it has allocated to $r$ in the current HTVA which $A$ is maintaining.

\subsection{Thread family management: changing workload fraction}
An important aspect of thread management is thread shrinking, that is the transition to a phase where a thread is allocated a lower percentage of an agent's resources (primarily time) than before. Thread shrinking involves a reduction of its intended workload. These allocations are intentions rather than allocations. Thus each thread is allocated a fraction of each of the available resources and that fraction counts as an expectation value for the relevant future. The length of that future may depend on the thread.

Planning may lead to a succession of expected resource allocations over an extended period of time.

Complementary to thread shrinking is thread growth, both being conditional upon the other in almost circular ways. I propose to view active thread shrinking as the meta-action which takes priority an which must precede the growth of another thread. In the presence of an executive thread capacity (i.e. workload fraction)  freed by thread shrinking is temporarily allocated to the executive thread in advance of its reallocation to another thread which either may be newly initiated, or reactivated, or just grown.

Thread shrinking for thread say $t$ may be performed while a different thread $r$ is in contemplating phase. In the absence of an executive thread this particular form of contemplation needs to take several other threads into account  in order to have $t$ (that is meta-actions performed within $t$) extend the resource allocation(s) for one or more other thread(s), say $s_1,..,s_n$, thus growing each of these so that in combination this compensates for the workload reduction that was imposed $r$, in preparation of making a C-switch to $r$ followed by an A-switch to $r$.

Remarkably in order to protect itself against the worry that thread shrinking on $t$ reduces the likelihood of its objectives being achieved it may be fruitful that $A$ becomes consciously proud of what has been achieved in $t$ thus far. Well-designed pride may endure in time and may facilitate subsequent thread growth in advance of 
resumption of its activities at a later stage.

\section{How to apply PMTh}
Applications of PMTh, if any, may come in various forms. I have one specific 
scenario in mind which motivates this work, and which I will briefly describe in this Section.

\subsection{Satisfycing applications count as real}
Suppose that (i) consultant $C$ brings agent $A$ into a state where $A$ is subjectively 
self-organized by way  of SIP driven multi-threading, (ii) $A$ is satisfied with this view of itself,
and (iii) $A$ actively maintains this view of itself in accordance with a workflow as suggested by $C$.

Now I claim that PMTh has found an application embodied in $A$'s novel HTVA architecture of itself. The
existence of an application of PMTh applies also if it has not been objectively 
established that $A$ is performing in some sense better than would be the case 
if $A$ would instead be working on the basis 
of some alternative view concerning the  architecture of its personal dynamics. It suffices (for PMTh having been
applied by $C$ to $A$) that $A$ is satisfied with the specific HTVA view of itself. 

How often and under what circumstances that state of affairs can be 
achieved, and how $C$ has to operate in order to obtain that result is a matter that is amenable for
empirical research. That holds too for the effectiveness of $A$'s agency depending on it maintaining a HTVA 
view of itself.

\subsection{PMTh based consulting as a thought experiment}
Assuming that applications of PMTh exist, a hypothetical PMTh based consultant $C$ 
may consult a client $A$ about
how to make best use of a PMTh including how to obtain a useful HTVA for $A$. Whether or not, 
and if so how, within a given
legislation this hypothetical consultancy strategy can legally be turned into practice will depend on the details 
of that legislation. It may be the case that a licence that $C$ has obtained for maintaining a consulting practice 
permits an experiment with PMTh based consulting, it may also be the case that only after some evidence base has been gathered $C$ is permitted to apply PMTh based consulting.

\subsubsection{Research on PMTh based consulting}
If research is needed to establish an evidence base for PMTh based consulting, then it suffices in principle 
to position that work within some research community and to comply with ethical standards of that community. 
Getting such research funded may be too difficult and for that reason alternative options for validating 
PMTh based consulting may be sought, for instance what I will call loosely coupled consulting.

\subsubsection{Loosely coupled consulting}
I will assume that there are no methodological and  
moral obstacles for $C$ to experiment in real
life cases with PMTh modelling of $A$ as long as this has the full cooperation of $A$, and as long as $A$ may be considered to be in full control of itself. In other words there is merely a loose coupling between $C$ and $A$ with $C$ helping $A$ to apply PMTh based modelling to itself.

Evidently questions asked by $C$ as well as suggestions made by $C$ must remain safely within
legal and moral boundaries. To this end the production of written minutes  of interactions between $A$ and $C$,
which must be regularly validated by 
$A$ and by $C$.  is critical. $C$ must allow $A$ to use these protocols as evidence 
when $A$ feels a need to complain about how $C$ has been working, or when clients of $A$ express 
claims that they have had a disadvantage because $A$ has followed $C$'s suggestions.

It seems possible to experiment with PMTh based consulting 
under the umbrella of a fairly detailed agreement that safeguards both parties against undertaking activities that
might be ethically criticized in hindsight. An ind this way gradually an evidence base may be collected that
justifies more direct PMTh base interventions by $C$.

\subsection{Interaction free PMTh based consulting}
Business ethical issues concerning the use of PMTh in a premature stage 
are less prominent if $C$ merely provides publicly available 
documentation from which $A$ can bring about a HTVA of its own dynamics and a way to 
proceed on that basis. The latter limited form of consultancy may be termed interaction free consultancy.

The documentation may for instance
contain abstract and anonimysed descriptions of how other clients of $C$ have applied PMTh. If
sufficiently many PMTh patterns are available to $A$ may find its own way. Clearly a moral difficulty arrises if 
$C$ proves fake user cases of its achievements with PMTh based consulting.

This form of interaction free consulting may be extended with (that is may be assumed to include) 
an anonymous interaction feature where $A$ is enabled in an anonymous manner to
ask $C$ general questions about how to understand the documentation. This protocol guarantees that $C$ 
does not enter in any specific considerations 
about $A$'s preferred HTVA or about how $A$ might optimally work in compliance with PMTh using its own HTVA. 
Questions and answers are
considered confidential, as  a way of safeguarding $C$'s business interests,  but at the same time these interactions must be faithfully documented by both parties and 
may be brought forward when formal complaints issued by either $A$ or by $C$ are to be dealt with.

\subsection{Conjectural abilities in connection with PMTh}
In~\cite{??} it is claimed that a novel theory $T$ aiming at being applicable in some area should preferably
be equipped with so-called conjectural abilities that a person or team that has taken $T$ on board is supposed
(by way of conjecture) to have in connection with achieving such applications.

The story on PMT hardly  qualifies as a theory because it constitutes merely a terminology 
for describing human agent behaviour in terms of computer science based terminology. 
Nevertheless I will attempt to formulate conjectural abilities that might come with an awareness of the story of PMT as set out in the paper.
I prefer to formulate the conjectural abilities as those of a consultant $C$ consulting a client $A$.
This formulation includes as a special case $A$ consulting itself on the basis of PMTh.
\begin{enumerate}
\item The ability to assist $A$ with designing a HTVA for itself. 
Awareness of the need that the HTVA is stable,
awareness of the need to develop threads each of which generate an independent sense of identity.
\item The ability to turn loosely coupled objectives into sequential plans (condensation) and to 
specify such plans in terms of instruction sequences. 
\item The ability to assist $A$ in designing a SIP for for its near future.
\item  A mechanical understanding of the issues that may stand in the way of efficient thread switching. Ability to consult $A$ on how to perform successful switches and pseudo-switches.
\item The ability to support $A$ in resolving confusion between PMTh and PMTa, so that the difficulty of PMTh thread switches will not be underestimated and the difficulty of PMTa task switches, as well as the difficulty of realising session concurrency, will not be overestimated.
\item The ability to consult $A$ on design rules for its HTVA maintenance which allows the dynamic 
management of its HTVA, by way of thread vector growth, thread vector shrinkage, and subthread mobility between threads, in such a way that at any time $A$ ``feels at home'' with its current HTVA.
\end{enumerate}

\subsection{Concluding remarks}
The story on PMTh is intended to constitute an initial attempt to make available the terminology and structure of thread algebra based concurrency theory for a human agent. Although my primary objective is to develop intellectual  tools for a human agent, at the same time this work may be considered a part of computer science as it may
support the development of computer programming based concepts, such as multi-threading and strategic interleaving, in a way which promotes a better intuitive understanding of these mechanisms, which may prove
useful for the further development of system design methods based on such concepts.

\end{document}